\pdfoutput=1
\documentclass[]{pasj01}
\usepackage{ulem}
\usepackage{textcomp}
\usepackage{appendix}
\draft

\begin{document} 
\title{New insights in giant molecular cloud hosting S147/S153 complex: signatures of interacting clouds}
\author{Dhanya\altaffilmark{1}, J.~S., Dewangan\altaffilmark{2}, L.~K., Ojha\altaffilmark{3}, D.~K., and Mandal\altaffilmark{1}, S.}
\email{}
\altaffiltext{1}{Malaviya National Institute of Technology (MNIT), Jaipur - 302 017, Rajasthan, India.}
\altaffiltext{2}{Physical Research Laboratory, Navrangpura, Ahmedabad - 380 009, India.}
\altaffiltext{3}{Department of Astronomy and Astrophysics, Tata Institute of Fundamental Research, Homi Bhabha Road, Mumbai - 400 005, India.}
\email{ jsdhanya89@gmail.com }

\maketitle
\begin{abstract}
In order to understand the formation of massive OB stars, we report a multi-wavelength observational study of a giant molecular cloud 
hosting the S147/S153 complex (size $\sim$90 pc $\times$ 50 pc). 
The selected complex is located in the Perseus arm, and contains at least five H\,{\sc ii} regions (S147, S148, S149, S152, and S153) powered by massive 
OB stars having dynamical ages of $\sim$0.2 -- 0.6 Myr. The Canadian Galactic Plane Survey $^{12}$CO line data (beam size $\sim$100$^{''}$.4) trace the complex in a velocity range of [$-$59, $-$43] km s$^{-1}$, and also reveal the presence of two molecular cloud components 
around $-$54 and $-$49 km s$^{-1}$ in the direction of the complex. 
Signatures of the interaction/collision between these extended cloud components are investigated through their spatial and velocity connections. These outcomes suggest the collision of these molecular cloud components about 1.6 Myr ago. 
Based on the observed overlapping zones of the two clouds, the collision axis appears to be parallel to 
the line-of-sight. 
Deep near-infrared photometric analysis of point-like sources shows the distribution of infrared-excess sources 
in the direction of the overlapping zones of the molecular cloud components, where all the H\,{\sc ii} regions are also spatially located.
All elements put together, the birth of massive OB stars and embedded infrared-excess sources seems to be triggered by two colliding molecular clouds in the selected site.
High resolution observations of dense gas tracer will be required to further confirm the proposed scenario.
\end{abstract}
\section{Introduction}
\label{sec:intro}
H\,{\sc ii} regions are considered as the first observable manifestations of massive OB stars ($\gtrsim$ 8 M$_{\odot}$). 
Such stars inject a great amount of radiative and mechanical energy in their host galaxies during their entire lifespan, and significantly affect their surrounding environments (e.g., Deharveng et al. 2010). 
However, understanding the birth mechanism of massive OB stars is a long-standing issue in star formation research, which is still being debated \citep{Zinnecker2007,Tan2014}. There are three major theories of massive star formation available in the literature, which are ``Turbulent core accretion", ``Competitive accretion", and ``Cloud-Cloud Collision (CCC)". Over the last decade, to explain the formation of massive OB stars, 
the CCC process has gained significant attention against the 
other ones (e.g., Zinnecker et al. 2007; Tan et al. 2014; Fukui et al. 2018). 
Because, this particular process provides appropriate initial conditions for the formation of massive stars at the intersection of two molecular clouds (e.g., Habe \& Ohta 1992; Anathpindika 2010; Inoue \& Fukui 2013; Takahira et al. 2014, 2018; 
Haworth et al. 2015a,b; Wu et al. 2015, 2017a,b; Torii et al. 2017; Balfour et al. 2017; Bisbas et al. 2017, and references therein).

In this paper, we have selected an extended star-forming S147/S153 complex or a system of Galactic H\,{\sc ii} regions, which is 
embedded in a giant molecular cloud (GMC; M$_{cloud}$ $\sim$6.1 $\times$ 10$^{4}$ M$_{\odot}$; Tatematsu et al. 1985) located in the Perseus arm (see also Kahane et al. 1985). 
The GMC hosts two small groups of H\,{\sc ii} regions, which are powered by late O or early B types stars \citep{Crampton1978}. 
One group consists of the S152 and S153 H\,{\sc ii} regions, while the other one contains S147, S148, 
and S149 H\,{\sc ii} regions \citep{Tatematsu1985}. Using the $^{12}$CO line data (beam size $\sim$2$'$.7), they found two distinct cloud components at $-$50 and $-$54 km s$^{-1}$ in the direction of the GMC. 
The cloud at $-$50 km s$^{-1}$ (M$_{cloud}$ $\sim$1.9 $\times$ 10$^{4}$ M$_{\odot}$; see Table~1 in Tatematsu et al. 1985) hosts the S152 and S153 H\,{\sc ii} regions, while the S147, S148, and S149 H\,{\sc ii} regions are associated with the cloud at $-$54 km s$^{-1}$ (M$_{cloud}$ $\sim$4 $\times$ 10$^{3}$ M$_{\odot}$; see Table~1 in Tatematsu et al. 1985). 
Using the JCMT $^{12}$CO(J=2-1) line data (beam size $\sim$21$''$), \citet{Azimlu2011} also investigated two groups of molecular clumps toward the GMC. 
These authors found the molecular gas at [$-$51.8, $-$49.3] km s$^{-1}$ in the direction of the one group of molecular clumps, which are distributed toward the S152 and S153 H\,{\sc ii} regions. The other group of molecular clumps at [$-$55.1, $-$51.7] km s$^{-1}$ is seen toward the S147, S148, and S149 H\,{\sc ii} regions. 
However, there is no study available in the literature to examine the interaction of these two cloud components, 
which can enable us to infer the exact operational star-formation process in the S147/S153 complex. 
In the eastern side of the GMC, a supernova remnant SNR G109.1$-$1.0 (CTB 109) has been reported (see Figure~2a in Tatematsu et al. 1985 and also Gregory et al. 1980), and has a semi-circular shell morphology 
detected both in radio and X-ray data. \citet{Sofue1983} reported the distance and age of the SNR G109.1$-$1.0 to be 4.1 kpc and 1.5 $\times$ 10$^{4}$ yr, respectively (see also Tatematsu et al. 1985). \citet{Tatematsu1985} compared the age of SNR G109.1$-$1.0 with the ages of S152 and S153, which are much older in the order of 10$^{6}$ yr (see also their paper). Considering these observational facts, these authors ruled out the impact of the SNR 
on the surroundings of the selected complex, which is not explored in this paper. 
Earlier, \citet{Crampton1978} and \citet{Tatematsu1985} adopted a distance of $\sim$4 kpc
to the star-forming S147/S153 complex. In this paper, we have used the distance of 4.0 kpc for the S147/S153 complex. In the star-forming S147/S153 complex, the formation process of the powering OB stars of 
H\,{\sc ii} regions is unknown in the literature. To observationally explore the interaction of two cloud components present in the GMC and 
the birth process of massive OB stars, the Canadian Galactic Plane Survey (CGPS) $^{12}$CO (J=1$-$0) line data (1$\sigma$ $\sim$0.3 K km s$^{-1}$; Taylor et al. 2003) and the deep 
UKIDSS near-infrared (NIR) photometric data have been analyzed in the S147/S153 complex.    

The present paper is arranged as follows. The data sets used in this work are described in Section~\ref{sec:data}. 
Section~\ref{sec:res} presents new results obtained through a multi-wavelength approach.
A discussion on the observational findings is given in Section~\ref{sec:disc}.
Finally, we conclude in Section~\ref{sec:conc}.
\section{Data sets}
\label{sec:data}
Multi-wavelength data sets are adopted in this paper, and were retrieved from different existing surveys (see Table~\ref{tab1}). 
In the selected large-scale area toward the complex (size $\sim$90 pc $\times$ $\sim$50 pc; central coordinates: {\it l} = 108$^{\circ}$.515, {\it b} = $-$1$^{\circ}$.074), these data sets are utilized to explore the ionized gas, molecular gas, dust continuum emission, mid-infrared (MIR) emission, and infrared excess sources. 
\begin{table*}
\scriptsize
\centering
\tbl{A list of multi-wavelength surveys utilized in the present work.}{
\begin{tabular}{llll}
\hline 
  Survey  &  Wavelength/Frequency(s)       &  Resolution         &  Reference  \\   
\hline
\hline 
 NRAO VLA Sky Survey (NVSS)        & 1.4 GHz        & $\sim$45$\arcsec$          &\citet{Condon1998}\\
 {\it Planck} Survey               & 100~GHz, 353~GHz, 857~GHz                 & $\sim$9$'$.7, 4$'$.9, 4$'$.6          & \citet{planck14}\\
 Canadian Galactic Plane Survey (CGPS) $^{12}$CO(J=1-0)     &2.6 mm            & $\sim$100$\arcsec$.4         &\citet{Taylor2003}\\   
BOLOCAM Galactic Plane Survey (BGPS)      &1.1 mm         &$\sim$33$\arcsec$    &\citet{Aguirre2011}\\
Wide-field Infrared Survey Explorer (WISE)          & 12 $\mu$m        & $\sim$6$''$.5     & \citet{Wright2010}\\
UKIRT Near-Infrared Galactic Plane Survey (GPS)         &1.25 -- 2.2 $\mu$m                   &$\sim$0.8$\arcsec$     &\citet{Lawrence2007}\\
Two Micron All Sky Survey (2MASS)                                                 &1.25 -- 2.2 $\mu$m                  & $\sim$2$\arcsec$.5          &\citet{Skrutskie2006}\\
\hline          
\end{tabular}}
\label{tab1}
\end{table*}
\section{Results}
\label{sec:res}
\subsection{Physical environment of the S147/S153 complex}
\label{subsec:multi}
In this section, we present new observational outcomes derived using the analysis of the multi-wavelength data.
\subsubsection{Multi-wavelength view of the S147/S153 complex}
\label{subsec:aamulti}
\begin{figure*}
\begin{center}
\includegraphics[width=11cm]{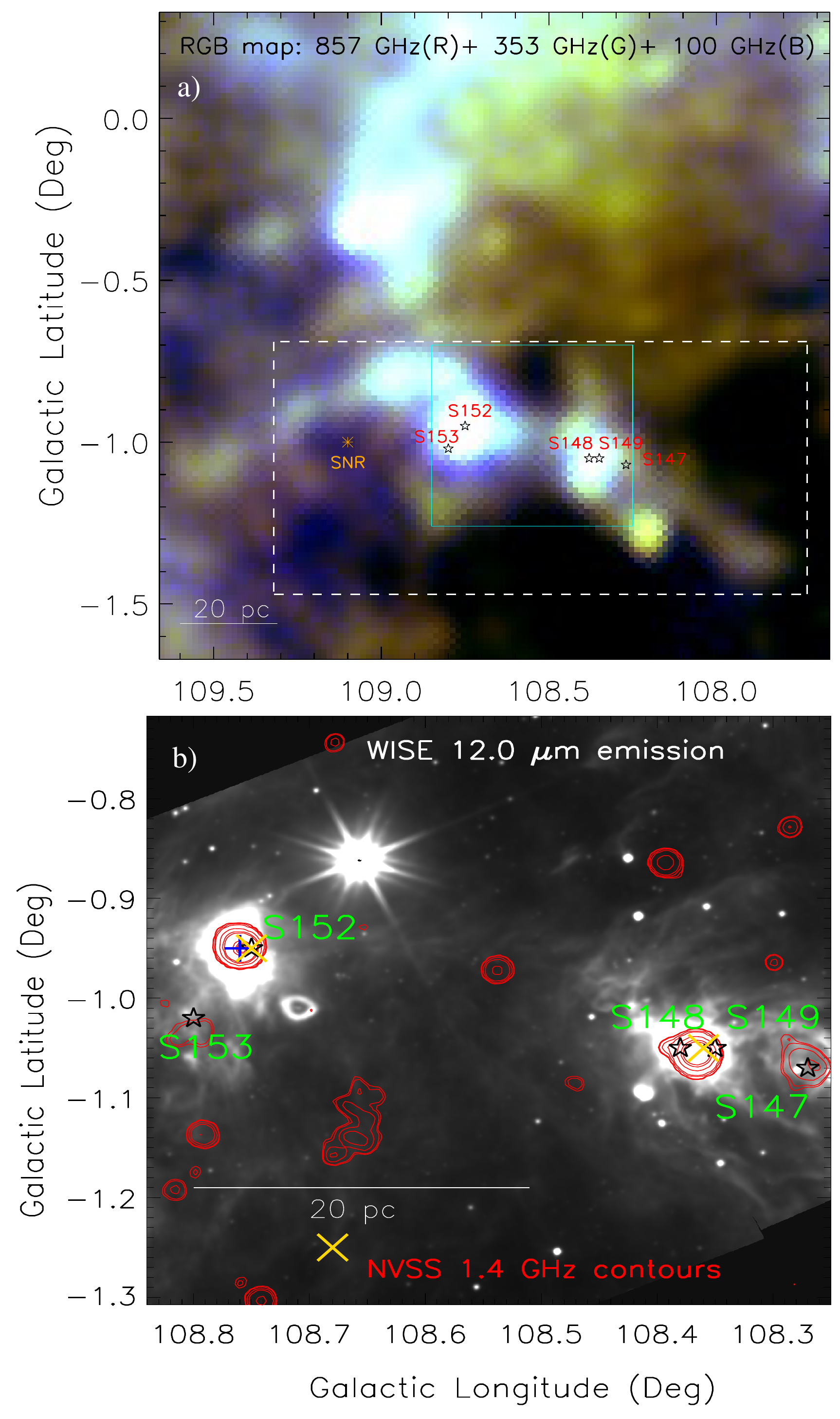} 
\end{center}
\caption{\scriptsize a) The {\it Planck} 3-color composite map of a large scale area 
containing the S147/S153 complex (size $\sim$2$^{\circ}$.0$ \times$ 2$^{\circ}$.0; central 
coordinates: {\it l} = 108$^{\circ}$.66, {\it b} = $-$0$^{\circ}$.68). The color-composite map (in log scale) is made using the {\it Planck} images at 857 GHz (red), 353 GHz (green), and 100 GHz (blue). A broken box (in white) indicates the area (size $\sim$90 pc $\times$ $\sim$50 pc; central coordinates: {\it l} = 108$^{\circ}$.515, {\it b} = $-$1$^{\circ}$.074), 
which is the target region of this paper. 
The position of SNR G109.1-1.0 is marked with an orange asterisk. 
b) The WISE 12 $\mu$m image of an area highlighted by a solid box (in cyan) in Figure~\ref{fig1}a. 
The image at 12 $\mu$m is also overlaid with the NVSS radio continuum emission contours at 1.4 GHz, which are shown with the 
levels of 2.18, 2.62, 4.37, 7.0, 26.20, 61.12, 87.32, 436.60, 611.24, 698.56, and 855.74 mJy/beam (where 1$\sigma$ = 0.45 mJy/beam; Condon et al. 1998). Multiplication symbols (in yellow) represent the locations of the 22 GHz water masers, while the position of the 6.7 GHz methanol maser is 
marked by a plus sign (in blue). In each panel, black stars indicate the positions of previously known S147, S148, S149, S152, and S153 H\,{\sc ii} regions. 
The scale bar corresponding to 20 pc (at a distance of 4.0 kpc) is shown in both the panels.}
\label{fig1}
\end{figure*}

Figure~\ref{fig1}a shows a color-composite map of a large scale view of the selected S147/S153 complex, which is made 
using three {\it Planck\footnote[1]{{\it Planck} (http://www.esa.int/Planck) is a project of the European Space Agency - ESA - with instruments provided by two scientific Consortia funded by ESA member states (in particular the lead countries: France and Italy) with contributions from NASA (USA), and telescope reflectors provided in a collaboration between ESA and a scientific Consortium led and funded by Denmark.}} images 
(857 GHz (red), 353 GHz (green), and 100 GHz (blue)). An area of the selected target complex is highlighted 
by a dashed box in Figure~\ref{fig1}a. In Figure~\ref{fig1}a, the positions of previously known five H\,{\sc ii} regions and SNR G109.1$-$1.0 (CTB 109) are also marked and labeled. One can also find two groups of H\,{\sc ii} regions in 
Figure~\ref{fig1}a (i.e., S152--S153, and S147--S148--S149). 
Note that the {\it Planck} images do not provide more insights into the H\,{\sc ii} regions due to their coarse 
resolutions (see Table~\ref{tab1}). 
Using the MIR image at WISE 12 $\mu$m, a zoomed-in view of the area containing five H\,{\sc ii} regions is displayed in Figure~\ref{fig1}b. 
The WISE image is also overlaid with the NVSS radio continuum emission contours at 1.4 GHz (where 1$\sigma$ = 0.45 mJy/beam; Condon et al. 1998), indicating the locations of the ionized regions in the complex. Some of the radio continuum emission peaks are associated with 
the extended MIR emission traced in the image at 12 $\mu$m. 

\begin{figure*}
 \begin{center}
  \includegraphics[width=11.4cm]{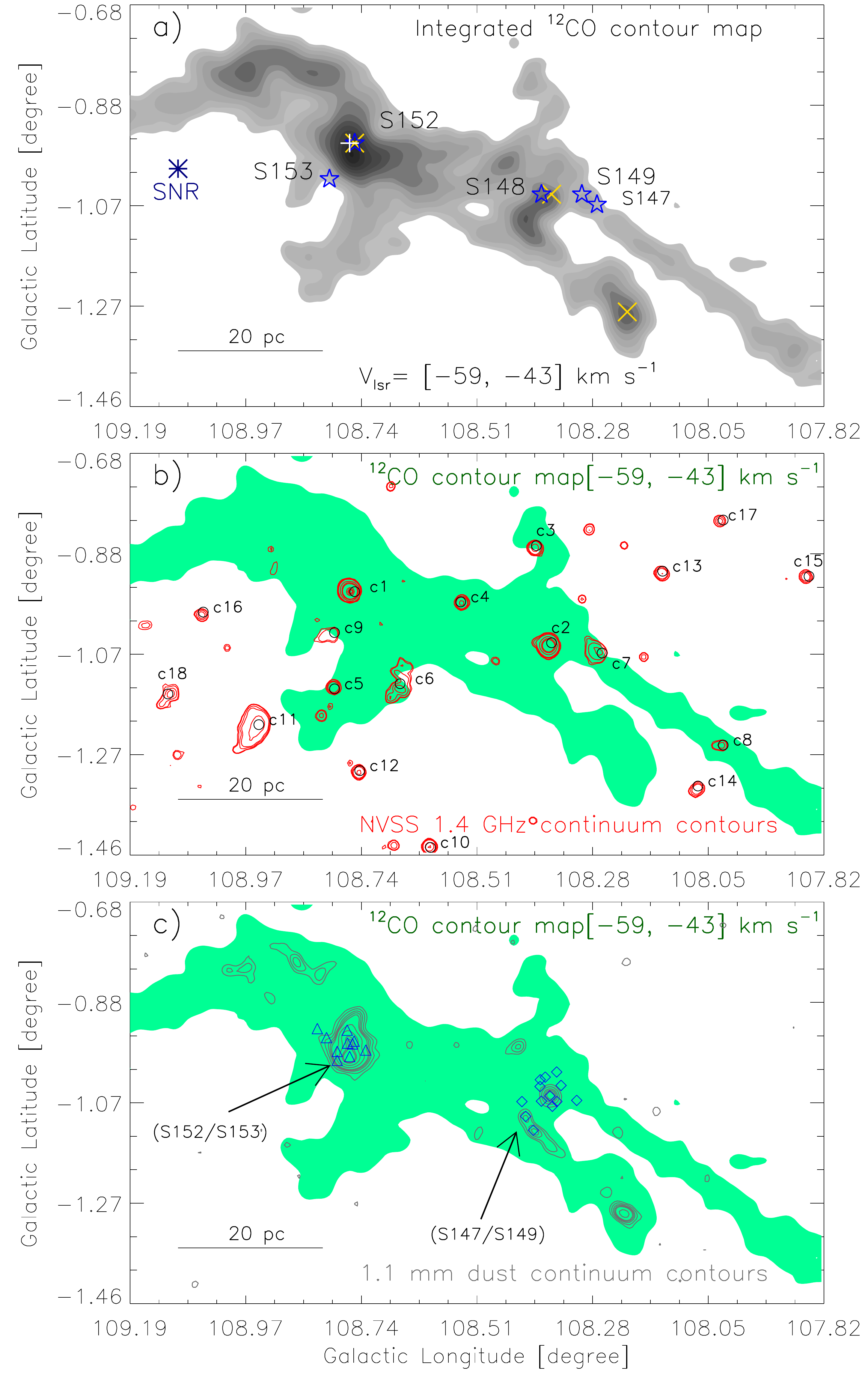} 
 \end{center}
\caption{\scriptsize a) The CGPS $^{12}$CO intensity contour map integrated over the velocity range of [$-$59, $-$43] km s$^{-1}$. 
The molecular emission contours are shown with the levels of 5.4, 7.9, 11.8, 17.8, 24.7, 29.6, 39.5, 49.3, 59.2, 69.1, 78.9, 88.8, and 97.7 K km s$^{-1}$ (where 1$\sigma$ $\sim$0.43 K km s$^{-1}$). 
The positions of SNR G109.1-1.0, S147, S148, S149, S152, S153, and masers are marked by the same symbols as shown in 
Figures~\ref{fig1}a and~\ref{fig1}b. Multiplication symbols (in yellow) represent the locations of the 22 GHz water masers, while the position of the 6.7 GHz methanol maser is marked by a plus sign (in white). 
b) Overlay of the NVSS 1.4 GHz radio continuum contours (in red) on the $^{12}$CO filled contour map. 
The molecular cloud boundary is highlighted by the $^{12}$CO contour with a level of 5.4 K km s$^{-1}$.
The NVSS 1.4 GHz contours are plotted with the same levels of as shown in Figures~\ref{fig1}b. 
Eighteen ionized clumps are marked by black circles along with corresponding IDs (see also Table~\ref{tab2}).
c) Overlay of the BOLOCAM 1.1 mm dust continuum contours (in black) on the $^{12}$CO filled contour map. 
The molecular map is the same as shown in Figure~\ref{fig2}b. 
The BOLOCAM 1.1 mm contours are shown with the levels of 3.26 Jy/beam $\times$ (0.014, 0.019, 0.035, 0.11, 0.15, 0.25, 0.8, 0.9), 
where 1$\sigma$ $\sim$11 mJy/beam \citep{Aguirre2011}. 
The positions of the $^{12}$CO molecular clumps (from Azimlu et al. 2011) are also marked in the 
panel (see diamonds and triangles).}
\label{fig2}
\end{figure*}

Figure~\ref{fig2}a shows an integrated intensity map of CGPS $^{12}$CO (J = 1--0) from $-$59 to $-$43 km s$^{-1}$, tracing the molecular cloud associated with the S147/S153 complex. The positions of five H\,{\sc ii} regions, 22 GHz water masers 
(V$_{lsr}$ $\sim$$-$45 -- $-$52 km s$^{-1}$; Sunada et al. 2007), SNR G109.1$-$1.0, and 6.7 GHz methanol maser (V$_{lsr}$ $\sim$$-$45.7 km s$^{-1}$; Xu et al. 2003) are also highlighted in Figure~\ref{fig2}a. 
In general, the 6.7 GHz methanol maser is known as a good tracer of massive star formation \citep{walsh98,urquhart13}, and 
the water maser emission has been reported as an important signpost of star formation activity (Litvak 1969; Genzel \& Downes 1977; Greenhill et al. 1998). 
Figure~\ref{fig2}b displays the overlay of the NVSS radio continuum emission contours at 1.4 GHz on the molecular map.
Several ionized clumps are seen within the molecular cloud associated with the selected complex (see Section~\ref{subsec:radio} for quantitative estimates). In Figure~\ref{fig2}c, we have superimposed the BOLOCAM dust continuum emission contours at 1.1 mm on the molecular map. 
Using the BOLOCAM dust continuum data, we have qualitative found the locations of dense clumps in the direction of the molecular cloud associated with the selected complex (see Figure~\ref{fig2}c). Using the JCMT $^{12}$CO(J=2-1) line data (beam size 21$''$), \citet{Azimlu2011} reported the positions of several molecular clumps in the direction of our target field, which are also marked in Figure~\ref{fig2}c (see diamonds and triangles). 

Together, Figure~\ref{fig2} provides the locations of the H\,{\sc ii} regions, dust clumps, and signposts of star formation in the direction of the selected GMC. 
\subsubsection{Ionized clumps in the S147/S153 complex}
\label{subsec:radio}
Based on the NVSS radio continuum map, we have identified the ionized clumps in the selected complex.
In the NVSS map at 1.4 GHz, we employed the {\it clumpfind} IDL program \citep{williams94} to depict the ionized clumps 
and to estimate their integrated flux densities. Several radio continuum contour levels were given as an input parameter for the 
{\it clumpfind}, and the value of the lowest contour level was about 5$\sigma$ (where 1$\sigma$ = 0.45 mJy/beam; Condon et al. 1998; see NVSS contours in Figure~\ref{fig1}b). Eighteen ionized clumps (i.e., c1--c18) are obtained in the complex, and are highlighted and labeled in Figure~\ref{fig2}b.  
In Figure~\ref{fig2}b, the radio clumps c1--c9 are seen in the direction of the boundary of the GMC hosting the S147/S153 complex, while other radio clumps (i.e., c10--c18) are away from the molecular cloud boundary. Among the radio clumps c1--c9, we find that the clumps (c1, c2, c7, and c9) are associated with the previously known H\,{\sc ii} regions. 
We have estimated the number of Lyman continuum photons (N$_{uv}$) for each ionized clump using the following equation \citep{matsakis76}:
\begin{equation}
N_{uv} (s^{-1}) = 7.5\, \times\, 10^{46}\, \left(\frac{S_{\nu}}{Jy}\right)\left(\frac{D}{kpc}\right)^{2} 
\left(\frac{T_{e}}{10^{4}K}\right)^{-0.45}\, \left(\frac{\nu}{GHz}\right)^{0.1}
\end{equation}
\noindent where S$_{\nu}$ is the measured total flux density in Jy, D is the distance in kpc, 
T$_{e}$ is the electron temperature, and $\nu$ is the frequency in GHz.  
The calculation uses the electron temperature of 10000~K and a distance of 4.0$\pm$0.5 kpc. The spectral type of the powering source of each ionized clump is derived by comparing the observed Lyman continuum flux against the theoretical value given in \citet{panagia73}. 
Table~\ref{tab2} contains the positions, radius (R$_{HII}$), integrated flux, Lyman continuum photons, and spectral type of each ionized clump.  
We find the presence of massive OB stars in our selected complex (see Table~\ref{tab2}). 
Using the values of N$_{uv}$ and R$_{HII}$, one can also compute the dynamical age (t$_{dyn}$) of each ionized clump 
using the following expression \citep{dyson80}: 
\begin{equation}
t_{dyn} = \left(\frac{4\,R_{s}}{7\,c_{s}}\right) \,\left[\left(\frac{R_{HII}}{R_{s}}\right)^{7/4}- 1\right] 
\end{equation}
where c$_{s}$ is the isothermal sound velocity in the ionized gas (c$_{s}$ = 11 km s$^{-1}$; \citet{bisbas09}), 
R$_{HII}$ (with 20\% uncertainty) is previously defined, and R$_{s}$ is the radius of the Str\"{o}mgren sphere (= (3 N$_{uv}$/4$\pi n^2_{\rm{0}} \alpha_{B}$)$^{1/3}$, where 
the radiative recombination coefficient $\alpha_{B}$ =  2.6 $\times$ 10$^{-13}$ (10$^{4}$ K/T)$^{0.7}$ cm$^{3}$ s$^{-1}$ \citep{kwan97}, 
``n$_{0}$'' is the initial particle number density of the ambient neutral gas, and N$_{uv}$ is defined earlier. 
Taking into account a typical value of n$_{0}$ (= 10$^{3}$ cm$^{-3}$), dynamical ages of the ionized clumps vary between $\sim$0.2 and $\sim$0.6 Myr. The values of the dynamical ages with estimated uncertainties are tabulated in Table~\ref{tab2}. 
Note that the calculation assumes that the H\,{\sc ii} region is homogeneous as well as spherically symmetric. 
Based on the analysis of the radio continuum data, one can also notice that many of the ionized clumps/H\,{\sc ii} regions have similar ages, which are distributed over $\sim$50 pc. It may indicate the onset of some triggering events for the star formation.
\subsection{Molecular cloud components in the S147/S153 complex}
\label{subsec:cogas}
In Figure~\ref{fig2}c, the boundary of the GMC hosting the S147/S153 complex is shown, where the locations of the previously observed 
molecular clumps (from Azimlu et al. 2011) are also indicated (see diamonds and triangles).
Blue triangles show the molecular clumps at [$-$51.8, $-$49.3] km s$^{-1}$ (having effective radius range $\sim$0.25--0.42 pc) toward the S152 and S153 H\,{\sc ii} regions, while blue diamonds indicate the molecular clumps at [$-$55.1, $-$51.7] km s$^{-1}$ (having effective radius range $\sim$0.51--1.07 pc) in the direction of the S147, S148, and S149 H\,{\sc ii} regions. 
We also examined the distribution of radial velocities of all these molecular clumps against their Galactic 
latitude positions, and found a velocity connection of these two groups of the molecular clumps (or molecular cloud components) 
in the GMC (not shown here). 
In order to further examine this result, we have produced the $^{12}$CO profiles, first moment map, second moment map, velocity channel maps, and position-velocity maps of the CGPS $^{12}$CO gas.

\begin{figure*}
\centering
\includegraphics[width=14 cm]{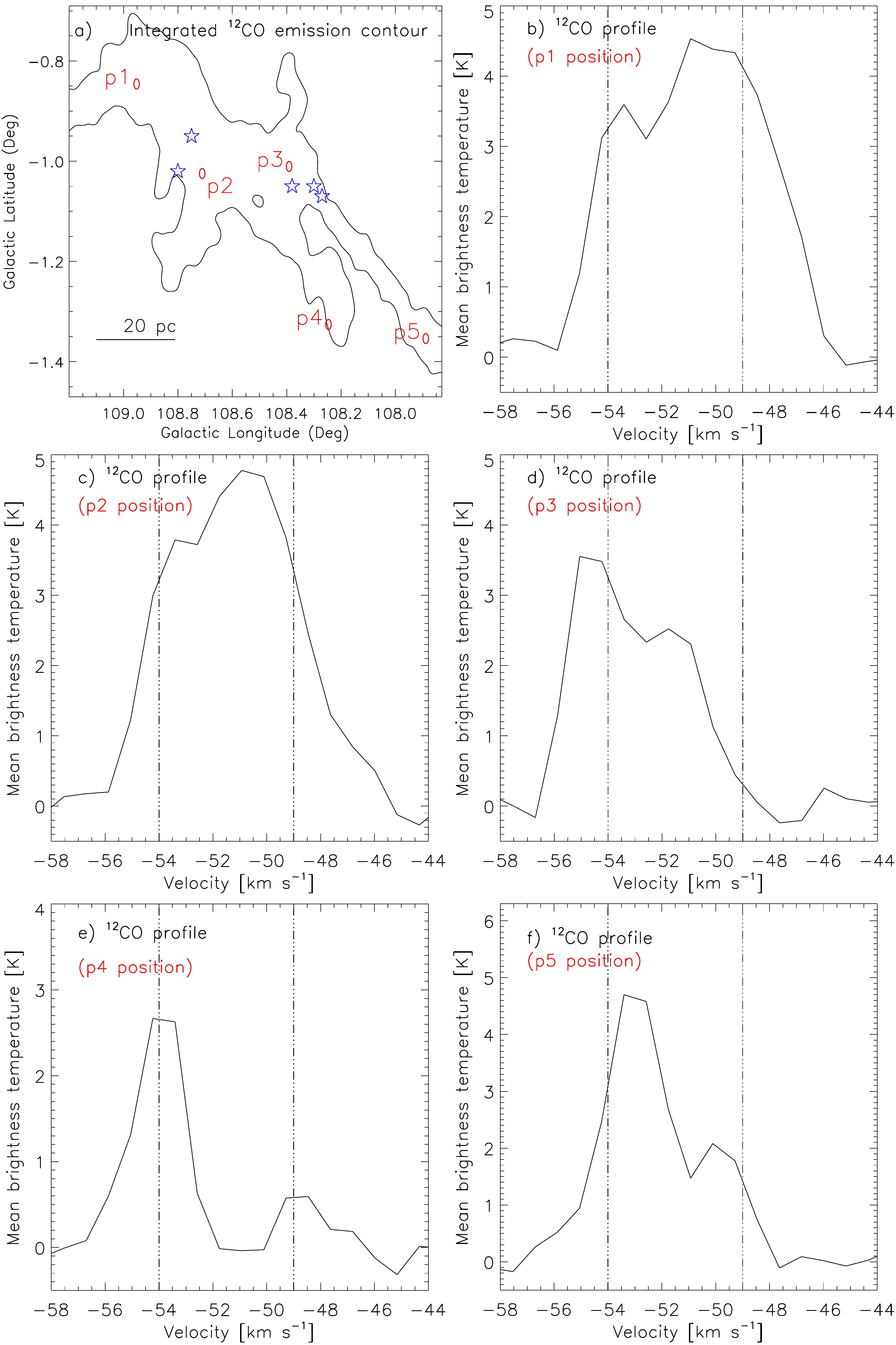} 
\caption{\scriptsize a) The molecular cloud boundary is highlighted by the $^{12}$CO contour with a level of 5.4 K km s$^{-1}$. Stars (in blue) indicate the positions of previously known S147, S148, S149, S152, and S153 H\,{\sc ii} regions. Additionally, five positions (i.e., p1 to p5) are also indicated by small ellipses in the figure.  
b--f) The $^{12}$CO profiles in the direction of five positions (i.e., p1 to p5; see corresponding positions in Figure~\ref{fig4y}a). Two velocities at $-$54 and $-$49 km s$^{-1}$ are indicated by two vertical broken lines.}
\label{fig4y}
\end{figure*}

\begin{figure*}
\centering
\includegraphics[width=14.7 cm]{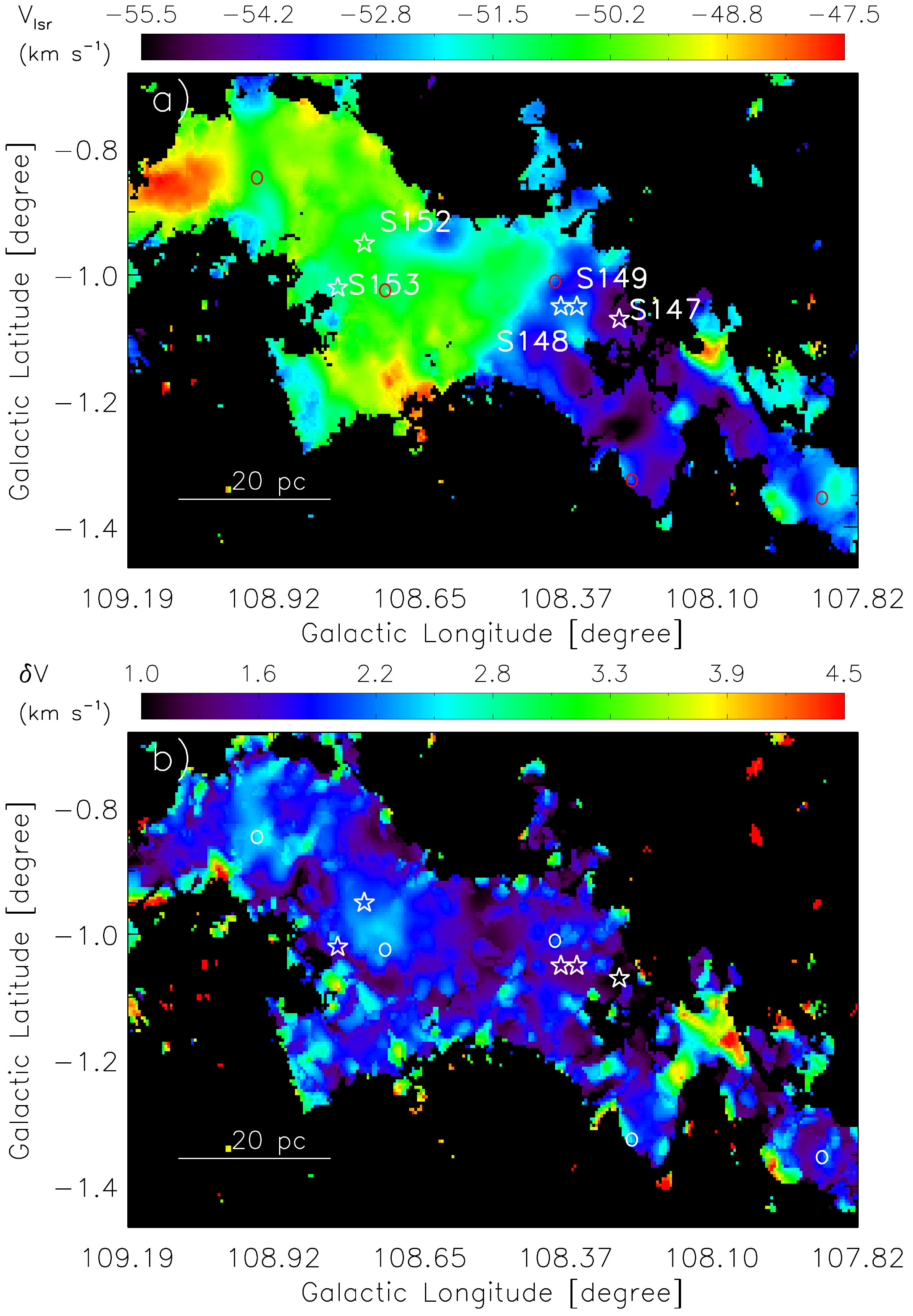}
\caption{\scriptsize a) Intensity-weighted mean velocity map (or the first moment map) of the CGPS $^{12}$CO. 
The color bar indicates the mean V$_{lsr}$ (in km s$^{-1}$). b) Intensity-weighted dispersion map (or the 
second moment map) of the CGPS $^{12}$CO. The color bar indicates the velocity dispersion (in km s$^{-1}$). In each panel, stars indicate the positions of the S147, S148, S149, S152, and S153 H\,{\sc ii} regions. 
In all the panels, open circles show the positions, where two velocity peaks are seen (see Figure~\ref{fig4y}).}
\label{fig4}
\end{figure*}

In Figure~\ref{fig4y}a, we have marked five positions (i.e., p1--p5; see red ellipses) in the direction of the boundary of the GMC hosting the S147/S153 complex. Figures~\ref{fig4y}b--\ref{fig4y}f show the observed $^{12}$CO profiles for different 
positions (i.e., p1--p5) marked in 
Figure~\ref{fig4y}a, revealing two velocity peaks (around $-$54 and $-$49 km s$^{-1}$; see broken lines in each panel). The spectra are obtained by averaging the region highlighted by an ellipse in Figure~\ref{fig4y}a. 
In Figure~\ref{fig4}a, we present the first moment map of $^{12}$CO, depicting the intensity-weighted mean velocity of the emitting gas. 
Considering the radial velocity observed toward the S152 H\,{\sc ii} region against the S148 H\,{\sc ii} region (see radial velocity bar in Figure~\ref{fig4}a), velocity gradient is clearly seen in the first moment map, suggesting the presence of two velocity components (see also position-velocity maps in this paper). Figure~\ref{fig4}b displays the second moment map (or intensity-weighted dispersion map), tracing a high velocity dispersion 
value ($>$ 1.6 km s$^{-1}$) toward the GMC hosting the S147/S153 complex. In the first and second moment maps, we have also highlighted the positions, where double velocity peaks are observed (see open circles in Figures~\ref{fig4y}a and~\ref{fig4y}b).

\begin{figure*}
\centering
\includegraphics[width=17 cm]{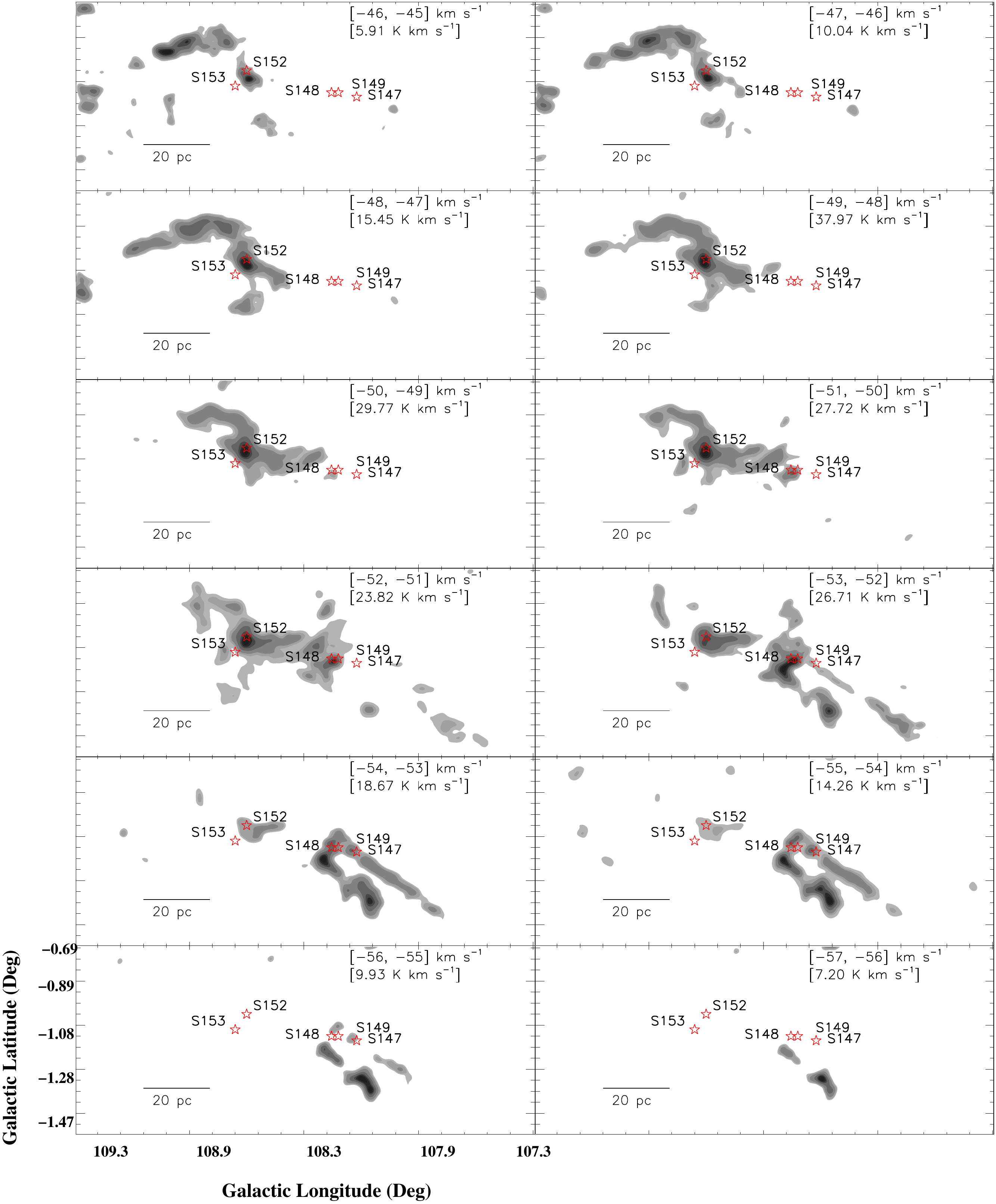}
\caption{\scriptsize Velocity channel maps of $^{12}$CO toward the selected GMC. 
The molecular emission is integrated with an interval of 1 km s$^{-1}$. 
The contour levels are (0.16, 0.2, 0.3, 0.5, 0.7, 0.8, 0.9) times of the peak value given in each panel in the unit of K km s$^{-1}$, where 1$\sigma$ $\sim$0.3 K km s$^{-1}$. Five H\,{\sc ii} regions (S147, S148, S149, S152, and S153) are also highlighted in all the panels.}
\label{fig5}
\end{figure*}

\begin{figure*}
\centering
\includegraphics[width=17.0 cm]{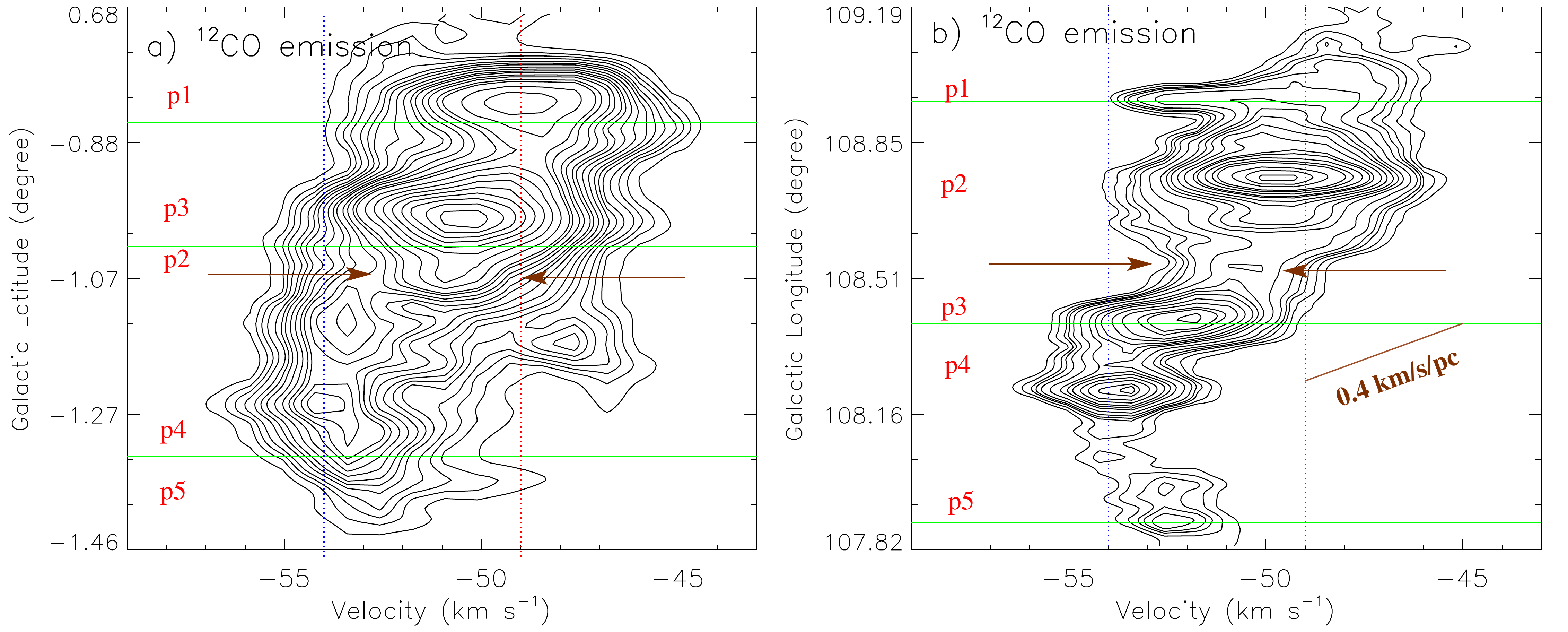}
\caption{\scriptsize a) Latitude-velocity map of $^{12}$CO. The molecular emission is integrated over the longitude range from 107$^{\circ}$.82 to 109$^{\circ}$.19. 
The contour levels are 30, 50, 65, 80, 100, 115, 130, 150, 170, 190, 205, 230, 250, 280, 300, 350, 400, 450, 500, 550, 600, 650, and 700 deg K, 
where 1$\sigma$ $\sim$5 deg K. 
b) Longitude-velocity map of $^{12}$CO. The molecular emission is integrated over the latitude range from -1$^{\circ}$.46 to -0$^{\circ}$.68.
The contour levels are 50, 65, 80, 90, 105, 130, 160, 180, 200, 230, 260, 300, 320, 360, 400, 430, 500, 540, and 580 deg K, where 1$\sigma$ $\sim$3.7 deg K. A scale bar referring to 0.4 km s$^{-1}$ pc$^{-1}$ is shown in the panel. In both the panels, green horizontal lines indicate the five positions (i.e., p1--p5; see Figure~\ref{fig4y}a), and 
two broken vertical lines indicate two cloud components (around $-$54 and $-$49 km s$^{-1}$). 
In each panel, two arrows (in brown) indicate an intermediate velocity emission.} 
\label{fig6}
\end{figure*}

Figure~\ref{fig5} displays the integrated CGPS $^{12}$CO velocity channel maps at intervals of 1 km s$^{-1}$.
The gas distribution in the channel maps suggests the presence of two clouds, and their spatial association 
(see panels at [$-$55, $-$54], [$-$53, $-$52], and [$-$49, $-$48] km s$^{-1}$ in Figure~\ref{fig5}). 
Figures~\ref{fig6}a and~\ref{fig6}b show the latitude-velocity and longitude-velocity maps, respectively. 
In Figure~\ref{fig6}b, we have also drawn a scale bar referring to 0.4 km s$^{-1}$ pc$^{-1}$, which can be used to examine the velocity gradient. The position-velocity maps of $^{12}$CO reveal two cloud components (around $-$54 and $-$49 km s$^{-1}$; see two broken vertical lines) 
in the direction of the complex. In the position-velocity maps, green horizontal lines show the five 
positions (i.e., p1--p5; see Figure~\ref{fig4y}a). Here, we again mention that the double peaks are found in the $^{12}$CO spectra toward these five positions (see Figure~\ref{fig4y}a).
In Figures~\ref{fig6}a and~\ref{fig6}b, two arrows (in brown) indicate a lower intensity intermediate velocity emission between two velocity peaks, showing a velocity connection of two cloud components. 

\begin{figure*}
\centering
\includegraphics[width=17.0 cm]{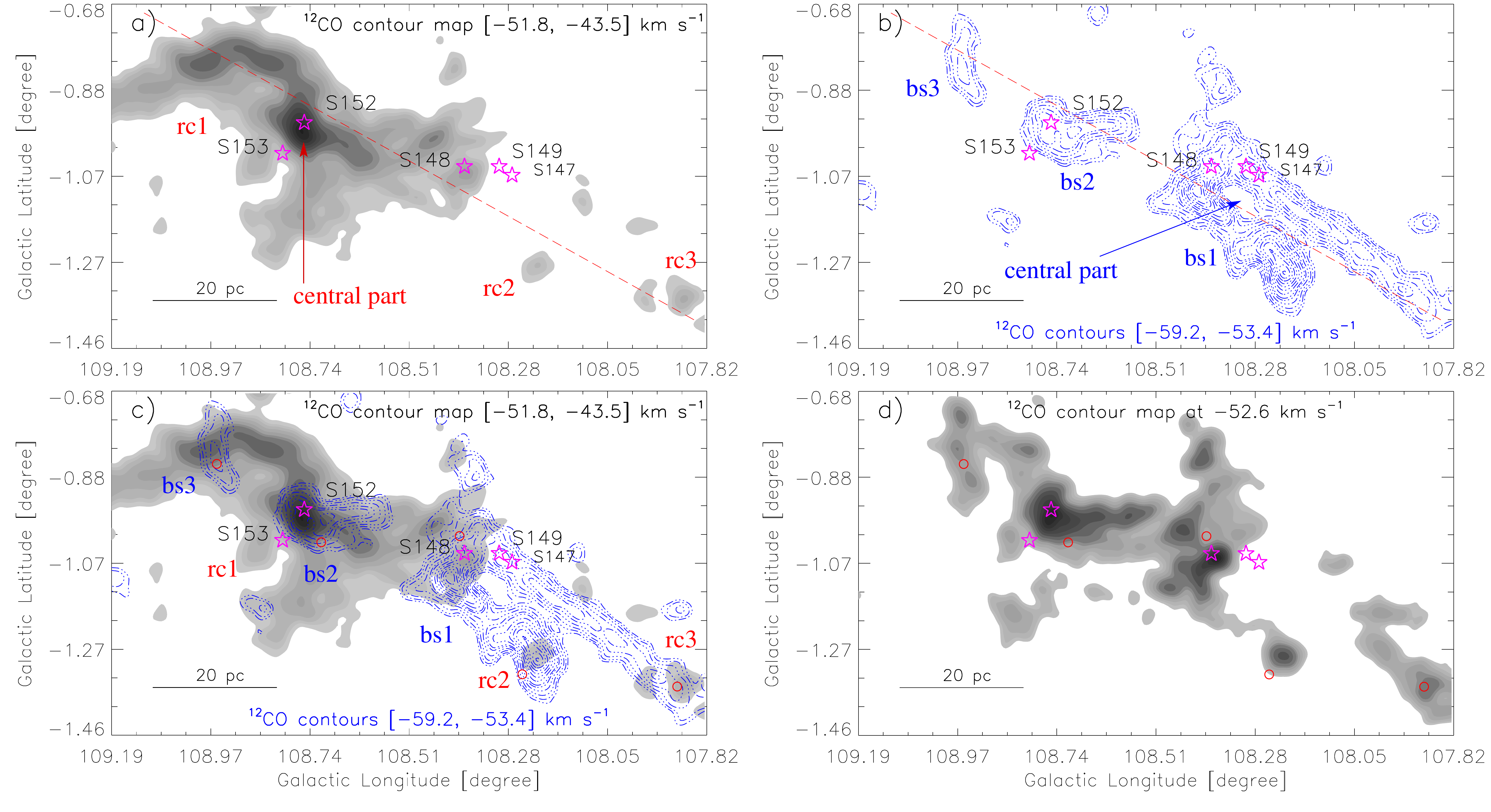} 
\caption{\scriptsize a) Spatial distribution of molecular gas associated with the cloud at [$-$51.8, $-$43.5] km s$^{-1}$ in the direction of the S147/S153 complex. The contour levels are 2.27, 4.16, 6.05, 9.07, 13.61, 18.90, 22.68, 
30.25, 37.81, 45.37, 52.93, 60.49, 68.05, and 74.86 K km s$^{-1}$, where 1$\sigma$ $\sim$0.42 K km s$^{-1}$.
Based on visual inspection, three parts of the cloud (i.e., ``rc1", ``rc2", and ``rc3") are labeled in the figure. 
An arrow indicates the central part of the cloud part ``rc1". 
b) Spatial distribution of molecular gas associated with the cloud at [$-$59.2, $-$53.4] km s$^{-1}$. 
The broken contours (in blue) are shown with the levels of 1.95, 2.84, 4.26, 6.39, 8.88, 10.66, 14.21, 
17.76, 21.31, 24.86, 28.41, 31.97, and 35.16 K km s$^{-1}$, where 1$\sigma$ $\sim$0.35 K km s$^{-1}$. 
On the basis of visual inspection, three parts of the cloud (i.e., ``bs1", ``bs2", and ``bs3") are labeled in the figure. 
An arrow highlights the central part of the cloud part ``bs1". 
c) Spatial distribution of molecular gas associated with the two clouds at [$-$59.2, $-$53.4] 
and [$-$51.8, $-$43.5] km s$^{-1}$ (see Figures~\ref{fig9}a and~\ref{fig9}b). 
d) Contour map of $^{12}$CO at $-$52.6 km s$^{-1}$ in the direction of the selected complex. 
The contour levels of the map are 0.72, 1.05, 1.58, 2.37, 3.29, 3.95, 5.27, 6.59, 7.9, 9.22, 10.54, 11.85, 13.04 K km s$^{-1}$, where 1$\sigma$ $\sim$0.3 K km s$^{-1}$. In the panels ``a" and ``b", an arbitrary broken line is marked, where selected cloud parts are almost aligned. 
In the panels ``c" and ``d", five circles show the positions, where the $^{12}$CO profiles are extracted (see Figure~\ref{fig4y}a). 
In each panel, the velocity information of the map is given, and the positions of S147, S148, S149, S152, and S153 H\,{\sc ii} regions are also highlighted by stars.}
\label{fig9}
\end{figure*}

Figure~\ref{fig9}a shows the spatial distribution of molecular gas associated with the cloud at 
[$-$51.8, $-$43.5] km s$^{-1}$ (or around $-$49 km s$^{-1}$). On the basis of visual inspection, at least three parts of this cloud component (i.e., ``rc1", ``rc2", and ``rc3") are marked in Figure~\ref{fig9}a, which appear to be spatially aligned (see a broken line in Figure~\ref{fig9}a). Another cloud component at [$-$59.2, $-$53.4] km s$^{-1}$ (or around $-$54 km s$^{-1}$) is presented in Figure~\ref{fig9}b. 
In Figure~\ref{fig9}b, based on visual inspection, we have also highlighted at least three parts of this 
cloud component (i.e., ``bs1", ``bs2", and ``bs3"). No molecular emission is seen toward the central area of the part ``bs1" (see an arrow in Figure~\ref{fig9}b), which may indicate the ionizing impact of massive stars associated with the S147, S148, and S149 H\,{\sc ii} regions. The part ``bs1" is also spatially aligned with the other two parts ``bs2" and ``bs3" (see a broken line in Figure~\ref{fig9}b). In Figure~\ref{fig9}c, we present the spatial distribution of these two molecular cloud 
components around $-$54 and $-$49 km s$^{-1}$. 
We find that the parts ``bs1", ``bs2", and ``bs3" are spatially overlapped with the cloud part ``rc1". 
Additionally, the cloud parts ``rc2" and ``rc3" are also spatially overlapped with the cloud part ``bs1". 
The five positions (i.e., p1--p5) are also indicated in Figure~\ref{fig9}c. 
Interestingly, these positions are seen toward the overlapping areas of these two cloud components. 
At least four H\,{\sc ii} regions associated with the sites S148, S149, 
S152, and S153 are also located toward the overlapping zones of the cloud components. 
Figure~\ref{fig9}d shows the contour map of $^{12}$CO at $-$52.6 km s$^{-1}$, which is an intermediate velocity value between two velocity components around $-$54 and $-$49 km s$^{-1}$. The map at $-$52.6 km s$^{-1}$ also traces compact molecular distribution toward the five positions. 

Altogether, the analysis of the CGPS $^{12}$CO line data (beam size $\sim$100$^{''}$.4) favours the spatial and velocity connections of two cloud components in the direction of the selected GMC. A detailed discussions on these findings are performed in Section~\ref{sec:disc}.
\subsection{Selection of infrared excess sources}
\label{subsec:Pop}
\begin{figure*}
\centering
\includegraphics[width=11.50 cm]{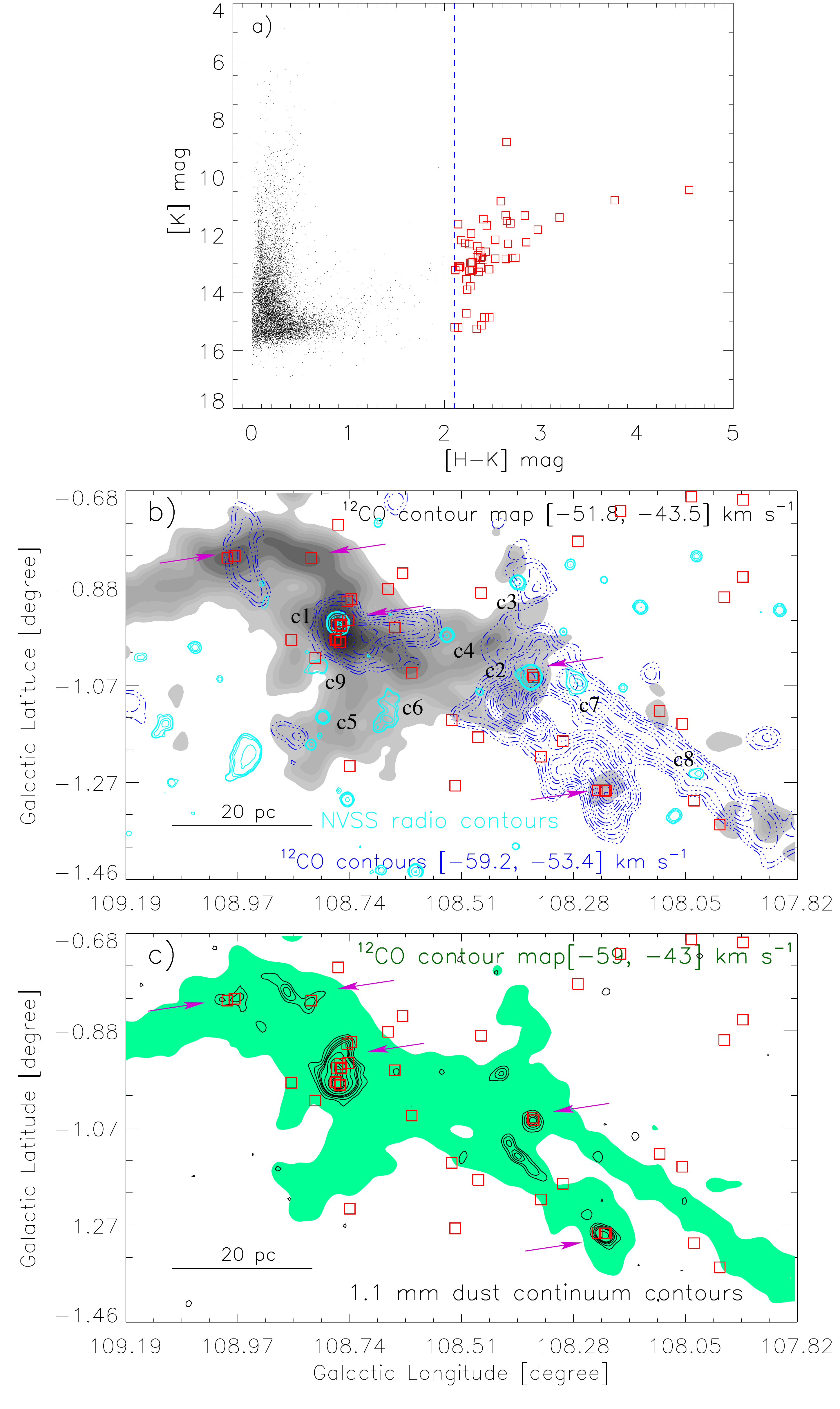}
\caption{\scriptsize a) NIR color-magnitude diagram (H$-$K vs K) of sources detected in the H and K bands.
These photometric magnitudes were obtained from the UKIDSS and 2MASS catalogs.
The selected infrared excess sources are highlighted by open red squares.
Black dots show stars with photospheric emission. 
Due to large numbers of these photospheric emission stars detected in the region, only some of them are displayed. 
The blue dashed line separates the selected infrared excess sources against other stars. b) Distribution of the selected infrared excess sources (see red squares) in the direction of two molecular cloud components (around $-$54 and $-$49 km s$^{-1}$). 
The $^{12}$CO maps are the same as shown in Figure~\ref{fig9}c. The NVSS 1.4 GHz radio continuum contours (in cyan) are also overlaid on the $^{12}$CO contour map (see also Figures~\ref{fig2}b). c) Overlay of the selected infrared excess sources (see red squares) on the integrated molecular map and the dust emission contours (see Figure~\ref{fig2}c). 
In the panels ``b" and ``c", arrows indicate the association of star formation activities with the dust clumps and/or the H\,{\sc ii} regions.}
\label{fig8}
\end{figure*}

To investigate the star formation activities in the S147/S153 complex, the knowledge of the infrared excess sources/young stellar objects (YSOs) in the selected area is needed. 
Hence, in this paper, the NIR color-magnitude scheme has been employed to identify a population of infrared excess sources. 
The excess emission of sources is expected due to the presence of circumstellar materials around them. Hence, these sources appear much redder in the NIR color-magnitude plot. 
In this work, the UKIDSS-GPS NIR data have been used for depicting more deeply embedded and faint young stellar populations, and are three magnitudes deeper than 2MASS. 
A reliable NIR photometric catalog of point-like sources was obtained from the UKIDSS GPS archival data release 
(see Dewangan et al. 2015 for more details, and also Lucas et al. 2008). 
Figure~\ref{fig8}a shows the NIR color-magnitude plot (H$-$K vs K) of sources that have detections in the H and K-bands. 
In Figure~\ref{fig8}a, the infrared excess sources with H$-$K $\gtrsim$ 2.1 mag are highlighted by squares. 
The color H$-$K cut-off (i.e., H$-$K $\sim$2.1 mag) is estimated by constructing the color-magnitude plot of sources detected in a nearby control-field region (central coordinates: {\it l} = 107$^{\circ}$.846, {\it b} = $-$1$^{\circ}$.88), which has the same size as that of the selected complex. The color-magnitude scheme yields a total of 105 infrared excess sources in our selected target field. 
In Figure~\ref{fig8}b, the positions of the infrared excess sources are overlaid on the two molecular maps 
around $-$54 and $-$49 km s$^{-1}$ (see also Figure~\ref{fig9}c). The NVSS radio continuum emission contours at 1.4 GHz are also shown in Figure~\ref{fig8}b. 
Three radio clumps (i.e., c1--c3) are found toward the overlapping zones of the two clouds. 
The infrared excess sources are distributed toward the radio clumps c1 (i.e., S152 H\,{\sc ii} region) and c2 (i.e., S148 H\,{\sc ii} region). 
The radio clump c1 is also associated with the 6.7-GHz methanol maser. In the direction of the molecular cloud associated with the complex, Figure~\ref{fig8}c shows the positions of the infrared excess sources against the dust continuum clumps at 1.1 mm. 
Figures~\ref{fig8}b and~\ref{fig8}c together suggest the association of star formation activities with the dust clumps 
and/or the H\,{\sc ii} regions, which are distributed toward the overlapping zones of the two clouds 
around $-$54 and $-$49 km s$^{-1}$ (see also arrows in Figures~\ref{fig8}b and~\ref{fig8}c). 
Furthermore, in the direction of the overlapping zones of the two clouds, the $^{12}$CO spectra show double peaks 
as well as broad profiles (see Figures~\ref{fig4y}b--\ref{fig4y}f). 
We find the absence of the wing features in the profiles, suggesting that the observed broad velocity widths in these regions may not be attributed to the outflows (see panels related to p1, p2, and p3 in Figure~\ref{fig4y}). Hence, the broad profiles appear to be observed due to the overlapping of the two velocity components along the line-of-sight.
\section{Discussion}
\label{sec:disc}
The selected GMC (M$_{cloud}$ $\sim$6.1 $\times$ 10$^{4}$ M$_{\odot}$; Tatematsu et al. 1985) in this paper has been known to 
host two groups of H\,{\sc ii} regions \citep{Tatematsu1985,Kahane1985}. In the direction of this molecular cloud, two velocity components/clouds (M$_{cloud}$ $\sim$0.4--1.9 $\times$ 10$^{4}$ M$_{\odot}$; Tatematsu et al. 1985) have also been reported in the literature (see also Azimlu et al. 2011). 
In the selected GMC, one group of molecular clumps at [$-$51.8, $-$49.3] km s$^{-1}$ is found toward the S152 and S153 H\,{\sc ii} regions, while the other group of molecular clumps at [$-$55.1, $-$51.7] km s$^{-1}$ is seen in 
the direction of the S147, S148, and S149 H\,{\sc ii} regions (see Azimlu et al. 2011 and 
also Figure~\ref{fig2}c in this paper). We have revisited the work of \citet{Azimlu2011}, and 
have found a connection of the two cloud components in the velocity (see Section~\ref{subsec:cogas}). 
This particular result indicates the possibility of the interaction or collision of these two 
cloud components, which is yet to be examined in the selected GMC. In this connection, we have carefully 
analyzed the tracers of the molecular gas, ionized gas, dust clumps, and embedded young stellar population toward our selected GMC.

As mentioned in the Introduction, the CCC process can produce massive stars at the junction of molecular clouds or 
the shock-compressed interface layer. In order to study the CCC process, \citet{habe92} carried out numerical simulations of head-on collisions of two non-identical clouds. They found gravitationally unstable cores/clumps at the interface of the 
clouds due to the effect of their compression, where massive stars can be triggered (e.g., Inoue \& Fukui 2013; Torii et al. 2017; Fukui et al. 2018). A spatial and velocity connection of two clouds is considered as one of the observational signposts of the CCC 
process (e.g., Torii et al. 2017; Dewangan 2017; Dewangan \& Ojha 2017; Dewangan et al. 2017,2018a,b,2019). The velocity connection of two clouds can be obtained via 
a lower intensity intermediate velocity emission (i.e., bridge feature) between them in the velocity space. The presence of the bridge feature might indicate the existence of a compressed layer of gas due to the collision of the clouds, which is also attributed to the turbulent layer created at the interface of the 
collision (e.g., Haworth et al. 2015a,b; Torii et al. 2017). 
Furthermore, the existence of a complementary spatial distribution of two clouds has been adopted as an another important signature of the CCC process (e.g., Torii et al. 2017; Fukui et al. 2018; Dewangan et al. 2018b,2019). 
In this context, one expects the spatial fit of ``Cavity/Keyhole/intensity-depression" and ``Key/intensity-enhancement" features.
In general, in the collision site, it is always not necessary to directly obtain the spatial fit of ``Key" and ``Key-hole" features. 
In such collision site, it is possible that one of these features could be displaced with respect to the other one 
(see Fukui et al. 2018, for more details). 
However, in a colliding system, a displacement may not be seen if the collision axis is parallel to the line-of-sight. 

Considering these highlighted observational signatures of the CCC process, we have carefully examined 
the CGPS $^{12}$CO line data in our selected target GMC. 
The position-velocity maps of $^{12}$CO reveal the presence of two molecular clouds 
at [$-$59.2, $-$53.4] km s$^{-1}$ (or around $-$54 km s$^{-1}$) and [$-$51.8, $-$43.5] km s$^{-1}$ (or around $-$49 km s$^{-1}$), which are also found to be linked by an intermediate velocity emission or a broad-bridge feature (see Section~\ref{subsec:cogas}). 
It implies the velocity connection of these two clouds. In Figure~\ref{fig9}c, we have displayed a spatial connection of the two cloud components toward our 
selected target GMC (see Section~\ref{subsec:cogas} for more details). The observed bridge feature in the velocity space may show spatially compact molecular distribution (see position-velocity maps), which seems to be consistent with the overlapping zones found in the selected GMC. 
These findings together favour a collision between the two cloud components in the past. 
Considering the observed overlapping zones of the two clouds (see Figure~\ref{fig9}c), the collision axis seems to be parallel to the line-of-sight. In this case, a displacement of one cloud with respect to other one may not be observed (e.g., Fukui et al. 2018). 
In the direction of the common zones of these two cloud components, 
the observed H\,{\sc ii} regions, a majority of infrared excess sources, the 6.7-GHz methanol maser, and dust clumps have been 
investigated (see Sections~\ref{subsec:cogas} and~\ref{subsec:Pop}). 
Hence, it may possible that the CCC process might have influenced the star formation history in the selected GMC. 
Therefore, we need to compute the collision timescale in the complex. 
The star formation activities are distributed upto about 20 pc (i.e., l$_{ccc}$) toward the overlapping zone of the clouds (see the area around S152/S153 in Figures~\ref{fig8}b and~\ref{fig8}c). 
With the help of the observed velocity separation range of the clouds (i.e., V$_{rel}$ = 5--12 km s$^{-1}$), 
the cloud collision timescale (l$_{ccc}$/V$_{rel}$) is estimated to be $\sim$3.9--1.6 Myr. 
In Section~\ref{subsec:cogas}, the dynamical ages of the H\,{\sc ii} regions in the complex have been found to be $\sim$0.2--0.6 Myr. 
It indicates that the collision timescale (i.e., $\sim$3.9--1.6 Myr) is older than the dynamical ages of the H\,{\sc ii} regions. 
However, the collision timescale (i.e. $\sim$1.6 Myr) corresponding to 12 km s$^{-1}$ appears more consistent with 
the dynamical ages of the H\,{\sc ii} regions and the ages of sites S152--S153 (i.e., about the order of 10$^{6}$ yr; Tatematsu et al. 1985). 
Additionally, a mean age of YSOs is reported to 
be $\sim$0.44--2 Myr \citep{evans09}. The 6.7-GHz methanol maser is also traced toward the S152 H\,{\sc ii} region, and suggests 
the existence of early stage of massive star formation ($<$ 0.1 Myr). 
In the simulations, \citet{inoue13} found that massive O stars can be produced within a few times of 10$^{5}$ yr. 
Considering these various age estimations and the observational signatures, it seems that the CCC process 
might have influenced the birth of massive OB stars and embedded YSOs in our selected GMC, which hosts the star-forming S147/S153 complex. This interpretation is still valid even if we assume about 10--20\% errors in the calculation of 
the collision timescale.
\section{Summary and Conclusions}
\label{sec:conc}
To study the formation process of massive OB stars in the GMC hosting the S147/S153 complex, an analysis of the multi-wavelength data has been performed in this paper.
The major outcomes of the paper are given below:\\
$\bullet$ Using the CGPS $^{12}$CO line data, the GMC (extension $\sim$90 pc) associated with the S147, S148, S149, S152, and S153 H\,{\sc ii} regions is studied in a velocity range of [$-$59, $-$43] km s$^{-1}$. These H\,{\sc ii} regions are excited by massive OB stars. 
The dynamical ages of these H\,{\sc ii} regions vary between $\sim$0.2 and $\sim$0.6 Myr for an ambient density of 10$^{3}$ cm$^{-3}$.\\  
$\bullet$ The CGPS $^{12}$CO line data reveal two molecular cloud components around $-$54 and $-$49 km s$^{-1}$ in 
the direction of the selected GMC. In the position-velocity space, these cloud components are connected through a broad-bridge feature at the intermediate velocity range. Spatial overlapping zones of these cloud components are investigated in the selected GMC.\\ 
$\bullet$ Using the deep UKIDSS and 2MASS NIR data sets, a total of 105 infrared-excess sources are 
selected in the target field.\\
$\bullet$ Signatures of the interaction between the cloud components are inferred through their spatial and velocity connections, which happened about 1.6 Myr ago. With the help of the observed overlapping zones of the two clouds, the collision axis is likely 
to be parallel to the line-of-sight. \\ 
$\bullet$ The powering massive OB stars of the H\,{\sc ii} regions, the 6.7-GHz methanol maser, and the embedded infrared excess sources are found at the 
overlapping zones of the two cloud components, where the dust continuum clumps at 1.1 mm are also observed. \\

Taking into account all the derived results in this paper, the formation of massive OB stars and embedded infrared excess sources seems to be triggered by two colliding molecular clouds in the S147/S153 complex. 
High resolution observations of dense gas tracer will be helpful to further confirm our interpretation.\\\\
%
We thank the anonymous reviewer for several useful comments and suggestions, which greatly improved the scientific 
contents of the paper. The research work at Physical Research Laboratory is funded by the 
Department of Space, Government of India.
This publication makes use of molecular line data from the Canadian Galactic Plane Survey (CGPS) carried out at the Dominion Radio Astrophysical Observatory (DRAO).
This work is based on data obtained as part of the UKIRT Infrared Deep Sky Survey.
This publication makes use of data products from the Two Micron All Sky Survey, which is a joint project of the University of Massachusetts and the Infrared Processing and Analysis Center/California Institute of Technology, funded by the National Aeronautics and Space Administration and the National Science Foundation.
\begin{table*}
\setlength{\tabcolsep}{0.3cm}
\centering
\tbl{Properties of the radio clumps derived using the NVSS 1.4 GHz radio continuum data (see Figure~\ref{fig2}b). The radio clumps c1--c9 are seen toward the boundary of the molecular cloud (see dagger symbols), and other radio clumps c10--c18 are located away from the cloud boundary (see Figure~\ref{fig2}b).}{
\begin{tabular}{lcccccccccl}
\hline 
  ID  &  l  &  b  &   R$_{HII}$     & S$_{\nu}$ & LogN$_{uv}$ & t$_{dyn}$    & Spectral Type & Remarks\\
    & (Deg) & (Deg)  & (pc)  & (Jy)  & (s$^{-1}$)  &  (Myr) & (dwarf main-sequence (V))&     \\     
\hline
\hline 
c1$\dagger$  & 108.75  & -0.95   & 0.37  & 1.35  & 48.23  & 0.21$\pm$ 0.01 &  O9--O9.5 & S152\\
c2$\dagger$  & 108.36  & -1.05   & 0.28  & 0.58  & 47.86  & 0.30$\pm$ 0.02 &  O9.5--B0 & S148/S149\\
c3$\dagger$  & 108.39  & -0.86   & 0.14  & 0.06  & 46.91  & 0.23$\pm$ 0.01 &  B0--B0.5 & -\\
c4$\dagger$  & 108.54  & -0.97   & 0.12  & 0.04  & 46.72  & 0.21$\pm$ 0.01 &  B0--B0.5 & -\\
c5$\dagger$  & 108.79  & -1.14   & 0.11  & 0.03  & 46.62  & 0.22$\pm$ 0.01 &  B0--B0.5 & -\\
c6$\dagger$  & 108.66  & -1.13   & 0.10  & 0.01  & 46.48  & 0.33$\pm$ 0.01 &  B0.5     & -\\
c7$\dagger$  & 108.26  & -1.07   & 0.11  & 0.03  & 46.65  & 0.51$\pm$ 0.01 &  B0--B0.5 & S147\\
c8$\dagger$  & 108.02  & -1.25   & 0.07  & 0.02  & 46.09  & 0.28$\pm$ 0.01 &  B0.5--B1 & -\\
c9$\dagger$  & 108.80  & -1.0	 & 0.08  & 0.01  & 46.26  & 0.42$\pm$ 0.01 &  B0.5--B1 & S153\\
c10          & 108.60  & -1.45   & 0.14  & 0.08  & 47.01  & 0.21$\pm$ 0.01 &  B0--B0.5 & -\\
c11          & 108.94  & -1.21   & 0.22  & 0.28  & 47.55  & 0.57$\pm$ 0.01 &  B0--B0.5 & -\\
c12          & 108.74  & -1.30   & 0.11  & 0.04  & 46.69  & 0.21$\pm$ 0.01 &  B0--B0.5 & -\\
c13          & 108.14  & -0.91   & 0.11  & 0.03  & 46.60  & 0.23$\pm$ 0.01 &  B0--B0.5 & -\\
c14          & 108.07  & -1.33   & 0.10  & 0.03  & 46.52  & 0.27$\pm$ 0.01 &  B0.5 & -\\
c15          & 107.85  & -0.92   & 0.09  & 0.02  & 46.37  & 0.22$\pm$ 0.01 &  B0.5--B1 & -\\
c16          & 109.05  & -0.99   & 0.08  & 0.01  & 46.26  & 0.23$\pm$ 0.01 &  B0.5--B1 & -\\
c17          & 108.02  & -0.81   & 0.08  & 0.01  & 46.20  & 0.21$\pm$ 0.01 &  B0.5--B1 & -\\
c18          & 109.11  & -1.14   & 0.11  & 0.02  & 46.59  & 0.47$\pm$ 0.02 &  B0--B0.5 & -\\
\hline           
\end{tabular}}
\label{tab2}
\end{table*}

\end{document}